\documentclass[runningheads]{llncs}

\usepackage{cite}
\usepackage{amsmath,amssymb,amsfonts}
\usepackage{algorithmic}
\usepackage{algorithm}
\usepackage{graphicx}
\usepackage{textcomp}
\usepackage[dvipsnames]{xcolor}
\usepackage{tabularray}
\usepackage{tabularx}
\usepackage{todonotes}
\usepackage[
urlcolor=blue,
]{hyperref}
\newcommand{\coo}{\ensuremath{\mathrm{CO_2}}}
\def\BibTeX{{\rm B\kern-.05em{\sc i\kern-.025em b}\kern-.08em
    T\kern-.1667em\lower.7ex\hbox{E}\kern-.125emX}}

\usepackage{fancyhdr}

\usepackage{listings}
\newcommand\YAMLcolonstyle{\color{black}\small\sffamily\mdseries}
\newcommand\YAMLkeystyle{\color{red}\small\sffamily\bfseries}
\newcommand\YAMLvaluestyle{\color{black}\small\sffamily\mdseries}

\definecolor{lightgray}{RGB}{240,240,240}

\makeatletter

\newcommand\language@yaml{yaml}

\expandafter\expandafter\expandafter\lstdefinelanguage
\expandafter{\language@yaml}
{
  keywords={true,false,null,y,n},
  keywordstyle=\color{darkgray}\bfseries,
  basicstyle=\YAMLkeystyle,                                 %
  sensitive=false,
  comment=[l]{\#},
  morecomment=[s]{/*}{*/},
  commentstyle=\color{purple}\ttfamily,
  stringstyle=\YAMLvaluestyle\ttfamily,
  backgroundcolor=\color{lightgray},
  moredelim=[l][\color{orange}]{\&},
  moredelim=[l][\color{magenta}]{*},
  moredelim=**[il][\YAMLcolonstyle{:}\YAMLvaluestyle]{:},   %
  morestring=[b]',
  morestring=[b]",
  literate =    {---}{{\ProcessThreeDashes}}3
                {>}{{\textcolor{red}\textgreater}}1     
                {|}{{\textcolor{red}\textbar}}1 
                {\ -\ }{{\mdseries\ -\ }}3,
}

\lst@AddToHook{EveryLine}{\ifx\lst@language\language@yaml\YAMLkeystyle\fi}
\makeatother

\newcommand\ProcessThreeDashes{\llap{\color{black}\mdseries-{-}-}}

\begin{document}

\title{Carbon-Awareness in CI/CD}

\author{
Henrik Claßen \and Jonas Thierfeldt \and Julian Tochman-Szewc \and \\
Philipp Wiesner \and Odej Kao
}
\authorrunning{Claßen et al.}
\institute{
Technical University Berlin, Berlin, Germany \\
\email{\{h.classen, thierfeldt, tochman-szewc\}@campus.tu-berlin.de} \\
\email{\{wiesner, odej.kao\}@tu-berlin.de}
}

\maketitle              %
\begin{abstract}

While the environmental impact of cloud computing is increasingly evident, the climate crisis has become a major issue for society.
For instance, data centers alone account for 2.7\% of Europe's energy consumption today.
A considerable part of this load is accounted for by cloud-based services for automated software development, such as continuous integration and delivery (CI/CD) workflows.

In this paper, we discuss opportunities and challenges for greening CI/CD services by better aligning their execution with the availability of low-carbon energy.
We propose a system architecture for carbon-aware CI/CD services, which uses historical runtime information and, optionally, user-provided information.
Our evaluation examines the potential effectiveness of different scheduling strategies using real carbon intensity data and 7,392 workflow executions of Github Actions, a popular CI/CD service.
Results show, that user-provided information on workflow deadlines can effectively improve carbon-aware scheduling.

\end{abstract}

\section{Introduction}

It is widely accepted that the main cause of anthropogenic climate is the release of greenhouse gases, such as carbon dioxide ($\coo$), into the atmosphere. %
The burning of fossil fuels such as coal and gas for energy production is a major contributor to the release of greenhouse gases. %
Globally, data centers, which are an important part of modern digital infrastructure, consumed about 205 TWh of electricity in 2018, representing about 1\% of global energy consumption~\cite{masanet_recalibrating_2020}, while in the EU they accounted for 2.7\% of energy consumption~\cite{energy-efficient}. Many cloud providers %
are taking steps to reduce the environmental impact of their data centers by utilizing more renewable energy.
However, the production of renewable energy sources such as solar and wind varies greatly over time and different locations, making it difficult to ensure a constant supply of energy.
A common metric to quantify the operational carbon emissions of energy consumption is called \emph{carbon intensity}, which describes the grams of $\coo$-equivalent greenhouse gases emitted per killowatt-hour of consumed energy (g$\coo$/kWh).

A significant portion of the work performed in cloud data centers pertains to services that enable the automated testing, building, and delivery of software, often referred to as continuous integration and delivery (CI/CD)~\cite{meyer_continuous_2014}.
According to a survey conducted by the CD Foundation\cite{cd_foundation_state_2022}, the use of CI/CD in the context of development and operation has become a standard practice today and has steadily increased. 
Out of 70,000 developers, about 47\% reported to use CI or CD in their software engineering process, as they increase productivity, software quality, and enable faster release cycles.

The operation of CI/CD services is a promising target for carbon-aware computing -- the alignment of a computing system's power usage with the availability of energy with low carbon intensity -- as they often include recurring or non-time-critical jobs.
Moreover, the execution of CI/CD workflows can be resource-intensive and thus energy-intensive, since separate virtual machines or containers are usually used for each individual job.
On the other hand, many CI/CD workflows are expected to execute as fast as possible and therefore have no temporal flexibility that can be leveraged.
In this paper
\begin{itemize}
    \item we discuss challenges and opportunities for carbon-aware CI/CD and propose a system architecture for more sustainable CI/CD services.
    \item we quantify potential improvements of carbon-aware scheduling using real-world carbon intensity and CI/CD workflow execution data.
\end{itemize}

The remainder of this paper is structured as follows.
Section~\ref{sec:relatedWork} reviews related work. 
Section~\ref{sec:discussion} discusses the main opportunities and challenges of carbon-awareness in CI/CD followed by a carbon-aware system architecture in Section~\ref{sec:architecture}.
Section~\ref{sec:evaluation} evaluates different scheduling scenarios.
Section~\ref{sec:conclusion} concludes the paper.

\section{Related Work}
\label{sec:relatedWork}

In this section we survey related works on CI/CD job management as well as general carbon-aware scheduling approaches.

\subsection{CI/CD Workflow Scheduling}

Major cloud and CI/CD service providers typically do not publicly disclose the specifics of their workflow scheduling heuristics. 
Jenkins, an open-source CI/CD automation server, employs a controller-agent architecture where a controller server schedules jobs to available agent nodes based on user-defined criteria, such as labels and priorities.
While cloud providers keep their scheduling algorithms under wraps, the research community has been advancing cloud computing scheduling algorithms.
For example, in the work of Ibrahim et al. \cite{ibrahim2020depth} different state of the art scheduling heuristics that aim to optimize resource allocation and load balancing in cloud environments were compared and evaluated. 
However, to the best of our knowledge, no service provider inherently includes features for carbon-aware job scheduling at this point.

\subsection{Carbon-Aware Computing}

Due to an increasingly critical public perception of unsustainable business practices and the fact that carbon pricing mechanisms, such as emission trading systems or carbon taxes, are being implemented on a global scale~\cite{WorldBank_CarbonPricing_2022}, the IT industry is actively striving to increase the utilization of low-carbon energy within datacenters. 
Carbon-aware computing aims to reduce the emissions linked to computing by adjusting flexible workloads in terms of both timing~\cite{wiesner_lets_2021, Radovanovic_Google_2021, Fridgen_NotAllDoomAndGloom_2021} and geographic locations~\cite{Zheng_MitigatingCurtailment_2020, Zhou_CarbonAwareLoadBalancing_2013, moghaddam_carbon-aware_2015} to align with clean energy sources.

Like the concept proposed in this paper, the majority of carbon-aware approaches aims at consuming cleaner energy from the public grid~\cite{hanafy2023carbonscaler, Radovanovic_Google_2021, Lin_CouplingDatacentersPowerGrids_2021, wiesner_lets_2021, Zhou_CarbonAwareLoadBalancing_2013}.
For example, Google defers delay-tolerant workloads during periods when power generation is associated with high carbon intensity~\cite{Radovanovic_Google_2021}, which describes the amount of greenhouse gas emissions per unit of consumed energy (gCO$_2$/kWh).
Although CI/CD workflows are sometimes used as exemplary use cases in evaluation scenarios~\cite{wiesner_lets_2021}, none of these works explicity adresses the oportunities and challenges arrising when implementing carbon-awareness in CI/CD services.

Besides optimizing for grid carbon intensity, recent works also try to better exploit renewable energy fluctuations directly.
For example, GreenSlot~\cite{goiri_matching_2015} is a scheduler that predicts the near-future availability of solar energy and schedules workload to maximize green energy consumption while also meeting job submission deadlines.
Similarly, Cucumber~\cite{Wiesner_Cucumber_2022} is an admission control policy which accepts low-priority workloads on underutilized infrastructure, only if they can be computed using excess energy.
Zheng et al.~\cite{Zheng_MitigatingCurtailment_2020} explore workload migration on underutilized data centers as a measure to reduce curtailment.

\section{Opportunities and Challenges}
\label{sec:discussion}
In this section, we discuss ways in which existing CI/CD services can be made more carbon-aware and debate what challenges arise in doing so.
After that, we look into the possibilities of utilizing additional user-supplied information to further improve carbon-aware scheduling.

\subsection{Leveraging Temporal Flexibility vs. \emph{Fail Fast}}

The \emph{fail fast} concept is a fundamental concept of CI/CD.
It involves providing fast feedback to developers on whether their code changes meet the requirements or not.
This helps to identify issues early in the development process, which can save time and resources in the long run. 

Many carbon-aware scheduling strategies involve shifting workloads to times when there is a greener energy mix or when there is renewable excess energy available. 
However, as the \emph{fail fast} concept demands workflows to be executed as soon and as fast as possible, it strongly conflicts with one of the most important properties of jobs suitable for carbon-aware computing: Temporal flexibility.
This conflict highlights the need for a careful analysis of which jobs are eligible and to what extent carbon-aware scheduling can be implemented without compromising the speed and efficiency of the development process.
Thus, a balance must be struck between these two concepts to achieve both fast feedback for developers and reduced carbon emissions.

\subsection{Promising CI/CD Workflows for Carbon-Aware Scheduling}

As stated above, not every CI/CD job is viable for full emission reducing scheduling, i. e. scheduling in time and location. Identifying this subset of jobs is the important first step in making CI/CD more carbon friendly. 

\begin{itemize}
    \item The first and most important category is made up of periodically running jobs. These are triggered by a date and time and not by a code change, such as nightly builds, integration test pipelines, or database backups.
    They are often not subject to the fail fast constraint: For example, the very requirement to perform a workflow \emph{nightly} is often to perform it \emph{outside of business hours}, which is usually more than half a day.
    The size of flexibility windows significantly impacts the potential for carbon footprint reduction~\cite{wiesner_lets_2021}.
    \item Second, CI/CD workflow often comprise of a multitude of interdependent jobs that are sometimes executed in parallel. Some of these steps may take multiple hours to complete, while other finish in a few minutes. Through the use of historical runtime data, carbon-aware schedulers can estimate flexibility windows for shorter tasks and leverage this information for carbon-aware scheduling.
    \item The third and last category is made up of jobs which are unnecessarily executed.
    Some code changes to a repository, e. g. changes to documentation, do not warrant the execution of, for example, unit tests. %
    Some research has already been done in this direction~\cite{abdalkareem_which_2021}.
    Detecting these cases and aborting the workflow yields the greatest benefit because their carbon emissions can be reduced by 100\%.
    However, identifying unnecessary CI/CD workflow executions is not the focus of this paper.
\end{itemize}

If there are multiple datacenters available to the CI/CD service to schedule workflows on, even workflows that cannot be shifted in time have a large potential in carbon reduction through the use of carbon-aware scheduling.
This potential, however, highly depends on data locality and data protection regulations of the specific workflow.

\subsection{Leveraging Historical Runtime Data}
Integrating carbon awareness into existing CI/CD solutions is the most sensible way to achieve a high level of adoption.
Therefore, we will now examine which information could be extracted from existing services without external input for carbon-aware scheduling.

The first bit of information needed is to identify a job's category (as defined above). 
Since a key concept of CI/CD services is automation, workflow definitions are seldom created on the fly but instead are defined in some machine readable format. 
These are then passed to the CI/CD framework used to execute. 
These configuration files, typically in YAML format, can also be fed into the scheduler to extract information about the parallelism, dependencies as well as periodicity of the jobs.
This information can then be used to try to determine the category and make decisions on the scheduling.
Additionaly, configuration files may contain further execution requirements  such as build system, mockup, or test data. 
As data transmission can also generate carbon emissions~\cite{ficher_assessing_2021}, it is furthermore important to inform the scheduler about where data is stored.

Data points produced by one or more actual executions of a job can also be used to influence scheduling decisions. 
For example, knowing the average and maximum runtime allows to better choose a time frame for execution.
Moreover, start and finish times of jobs can be used to guess dependencies between jobs which might have not been included in their definitions. 
The time frame between the end of one job and the start of the next earliest depending job offers flexibility for scheduling.
Even if a job is not dependent on another job, the start and end times can give an indication of a possible deadline:
Consider a job that is not triggered by some user activity, starting after office hours or in the night and finishing before the start of the next days office hours.
In this case, it can be assumed that this job is some kind of over night build or test and the time frame deducted from this can then be used for scheduling.
This is especially useful for jobs which run periodically.

The last source of information is the current state of the CI/CD system.
As service providers try to optimize their computing resources they cache container/VM images as well as required data at certain locations.
Furthermore, providers have to keep spare computing resources available at different sites to guarantee
fast response times (fail fast) under high system load.
A scheduler can make use of this information and the learned workflow execution requirements to influence where (machine, data center, or region) a job will be executed. 
For example, starting a new server creates higher associated carbon costs than consolidating work on running machines. 
As outlined before, moving data to a desired location also consumes energy and therefore produces carbon emissions. 
Following, if some or all data requirements for a job execution are present at a certain data center it can be more efficient to run the job there even if the carbon footprint of the electricity is higher than elsewhere. 

\subsection{Improving Scheduling with User-Supplied Information}
To further increase the effect of carbon-aware scheduling, we now investigate the potential effect of user-supplied information on the scheduling.
To implement this, a user interface must be provided by the CI/CD service provider.
An example is provided in Listing~\ref{lst:listing}.
Furthermore, users should be incentivized to provide additional information on their workflow runs and to accept potential delays in workflow executions.

One example of useful information that users can provide to the scheduler is runtime estimates.
Especially for new workflows where historical information is not available, the cold-start problem significantly hinders the scheduler's ability to perform sensible carbon-aware decisions.
If users are able to provide estimates of the duration of the job we can incorporating this information into the planning process to enable adjustments to the schedule, as needed.
However, it is important to consider the validity and accuracy of user information. This especially applies to the estimated runtime. It is known from high-performance computing that users more often than not tend to overestimate the runtime~\cite{cirne_comprehensive_2001, tsafrir_backfilling_2007, tang_fault-aware_2009, ward_scheduling_2002}. We expect the same to be true for CI/CD jobs. One possibility to counteract that is to gradually move to using historical data as it becomes available. The user should be informed of this change.

\begin{figure}[t]
\begin{lstlisting}[language=yaml, caption={Exemplary interface which allows users to provide additional information on a workflow's duration, deadline, and allowed regions for execution.}, label=lst:listing]
  ---
  name: CI/CD jobs
  on: [push]
  jobs:
    job-a:
      runs-on: ubuntu-latest
      carbon-aware: yes
      steps:
        - name: My first step
          uses: actions/hello_world@main
          with:
            duration: 1h
            deadline: 3h
            allowed-regions: [eu-central-1]
  ---
\end{lstlisting}
\end{figure}

Another highly valuable input for carbon-aware schedulers are deadlines.
As described above, if this information is not provided we can only try to guess a job's deadline based on interdependecies with other workflows. %
Since estimates must be conservative to avoid introducing delays or even errors, the potential carbon reductions can not be fully exploited.
With a user-specified deadline, this ambiguity is removed, jobs are still guaranteed to finish on time, and the full potential can be realized.
As described above, periodic jobs benefit the most from this.

Lastly, users should also be able to specify allowed regions for scheduling, for example, to avoid the scheduling in regions that require the migration of large volumes of data or that contradict with data protection regulations.
This information can is used to influence the selection of the assigned region, as shorter transfer times are taken into account.

\section{A Carbon-Aware CI/CD Service Architecture}
\label{sec:architecture}

\begin{figure*}[b!]
 \centering
 \includegraphics[clip, width=0.95\textwidth]{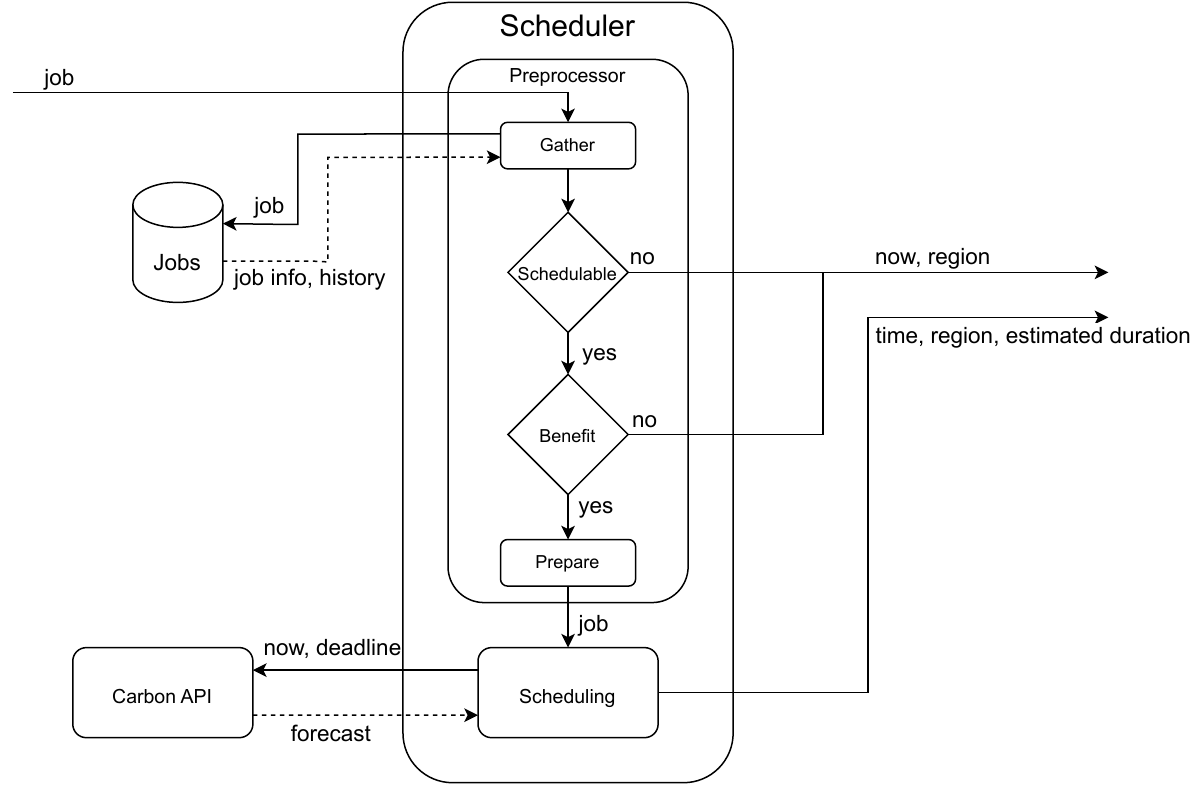}
 \caption{High-level systems architecture}
 \label{fig:architecture}
\end{figure*}

In the following, we present a high-level systems architecture for carbon-aware CI/CD services in a distributed computing environment.
We intentionally describe the concept in a a general manner and do not specify whether it is an internal part of a CI/CD framework or provided by a third party to ease the integration into existing solutions such as GitHub Actions, Jenkins, or proprietary services.
We assume that the scheduler is fed on a job-by-job basis and that it can query the information outlined in Section~\ref{sec:discussion} in an effective manner. 

To reduce the carbon footprint of CI/CD services, we first filter out promising jobs for carbon-aware scheduling and then plan their execution using information about previous runs, the state of the CI/CD system, and carbon intensity forecasts at different locations.
The overall concept is depicted in Figure~\ref{fig:architecture}.
When a CI/CD workflow starts, a request is made to the scheduler.
The result then contains when the job should be run, the region where to run and optionally the estimated duration that was used in the scheduling process.
If the job should run at a later point in time we need to store meta information about the job.

A preprocessor filters all incoming requests, to sort out jobs which cannot or should not be processed further. 
Jobs cannot be processed further if it is not possible to apply carbon-aware scheduling (see Section~\ref{sec:discussion}) and should not be processed further if the expected carbon savings are smaller than the costs of the scheduling.
Calculating this difference depends on many factors such as resource consumption of the scheduling process or length of the job. Therefore, we can not give a general answer when to apply the scheduling to a job and when not.
If a user provides an estimate for the runtime of a job, the duration used in the scheduling process is a combination of this estimate and historical data. The weight of the user input decreases with increasing historical data. Additionally, a time buffer is added to account for unforeseen increases in the duration, e. g. a sudden increase in the code base or tests. With an increasing number of previous runs the buffer is reduced.

The second step is the actual scheduling, which is the most important step, since the carbon reduction happens here. To estimate the carbon footprint, forecast data predicting the future carbon intensity per time and region is used. This data is fetched from an API for the time frame spanning the arrival time of the request to the (estimated) deadline of a job. An algorithm then uses this and available data of the CI/CD framework to estimate the total carbon footprint for each possible start time and region and selects the best result in terms of carbon intensity. For jobs that were previously filtered out, the currently best region in terms of carbon intensity is selected. To compare the prediction with the actual carbon emissions and provide feedback to the user, the runtime and energy consumption of a finished job is used to estimate the actual carbon intensity.

The interface for the human user to input information about a job (e. g. estimated runtime) can be implemented by a extending existing specification formats, see Listing~\ref{lst:listing}. This also applies to reporting results back to the user, by adding this data to CI/CD framework's reporting mechanisms.

\section{Evaluation of Carbon-Aware Scheduling Strategies}
\label{sec:evaluation}
We implemented a first prototype of our proposed architecture using Node.js and evaluate different carbon-aware scheduling strategies against it.
After describing the experimental setup, we present our results, and discuss the limitations of this analysis.

\subsection{Experimental Setup}

To model varying carbon intensity at different datacenters, we utilized marginal carbon intensity data provided by WattTime~\footnote{
\url{https://www.watttime.org}} over a four day period (October 12-16, 2022).
Marginal carbon intensity describes the carbon intensity of the power source that would meet any new demand -- also called marginal power plant.
For the example, the grid's marginal carbon intensity is very low if an additional kilowatt of demand would be produced by an otherwise shut-off wind turbine, but high, if it is generated by a gas turbine.
This makes marginal carbon intensity a very meaningful metric for scheduling decisions.

The data by WattTime contains actual and forecasted carbon intensity for 12 different regions across Europe, North America, and Australia.
Each region was assumed to have unlimited resource capacity for jobs to be executed.

To model the CI/CD service, we collected 7,392 historic workflow executions from the same time period by crawling ten popular GitHub repositories\footnote{
\href{https://github.com/alibaba/arthas}{alibaba/arthas}, \href{https://github.com/apache/dubbo}{apache/dubbo}, \href{https://github.com/apache/flink}{apache/flink}, \href{https://github.com/apache/spark}{apache/spark}, \href{https://github.com/k3s-io/k3s}{k3s-io/k3s}, 
\href{https://github.com/kubernetes/minikube}{kubernetes\-/minikube}$,\,$
\href{https://github.com/microsoft/typescript}{microsoft/typescript}$,\,$\href{https://github.com/microsoft/vscode}{microsoft/vscode}$,\,$\href{https://github.com/netdata/netdata}{netdata/netdata}$,\,$\href{https://github.com/Tencent/ncnn}{Tencent/ncnn}
} that perform their CI/CD using Github Actions.
The resulting dataset included the workflow name, start date, and runtime for each workflow execution.

We analyzed five experiments using different scheduling strategies.
For each experiment, we iterate over all jobs in the dataset in temporal order.
The five scheduling strategies are

\begin{itemize}
    \item \textbf{Round-robin.} This naive strategy uses round-robin scheduling establishing a baseline to assess the improvements introduced by carbon-awareness.
    \item \textbf{Location Shifting.} This strategy demonstrates the effectiveness of scheduling CI/CD jobs at locations of low carbon intensity.
    Yet unknown jobs — those for which we lack historical data to estimate the runtime — are scheduled using the round-robin method.
    \item \textbf{Location$\,+\,$Time Shifting (\{1, 3, 6\}h).} The remaining strategies show the full potential of carbon-aware scheduling by additionally enabling the scheduler to exploit temporal flexibility of jobs.
    This approach is tested across three cases with varying deadline buffers (1h, 3h, 6h) -- information which, in practice, could be provided by users.
    The job's deadline was set to the sum of the duration and the buffer, ensuring each job had the same flexibility window for rescheduling.
    We will abbreviate this strategy in the following section as LTS-\{1, 3, 6\}.
\end{itemize}

\subsection{Results}
\label{sec:results}
For each scenario, we report the distribution of jobs and accumulated carbon emissions over time relative to the round-robin baseline. We opted for relative results, as our dataset did not contain any energy consumption values and therefore we could not calculate carbon emissions in grams of CO$_{2}$eq or similar. This means that our findings provide an indication of the relative carbon intensity of different scheduling strategies, but not an absolute measure of carbon emissions.

\begin{figure*}
 \centering
 \includegraphics[clip, width=\textwidth]{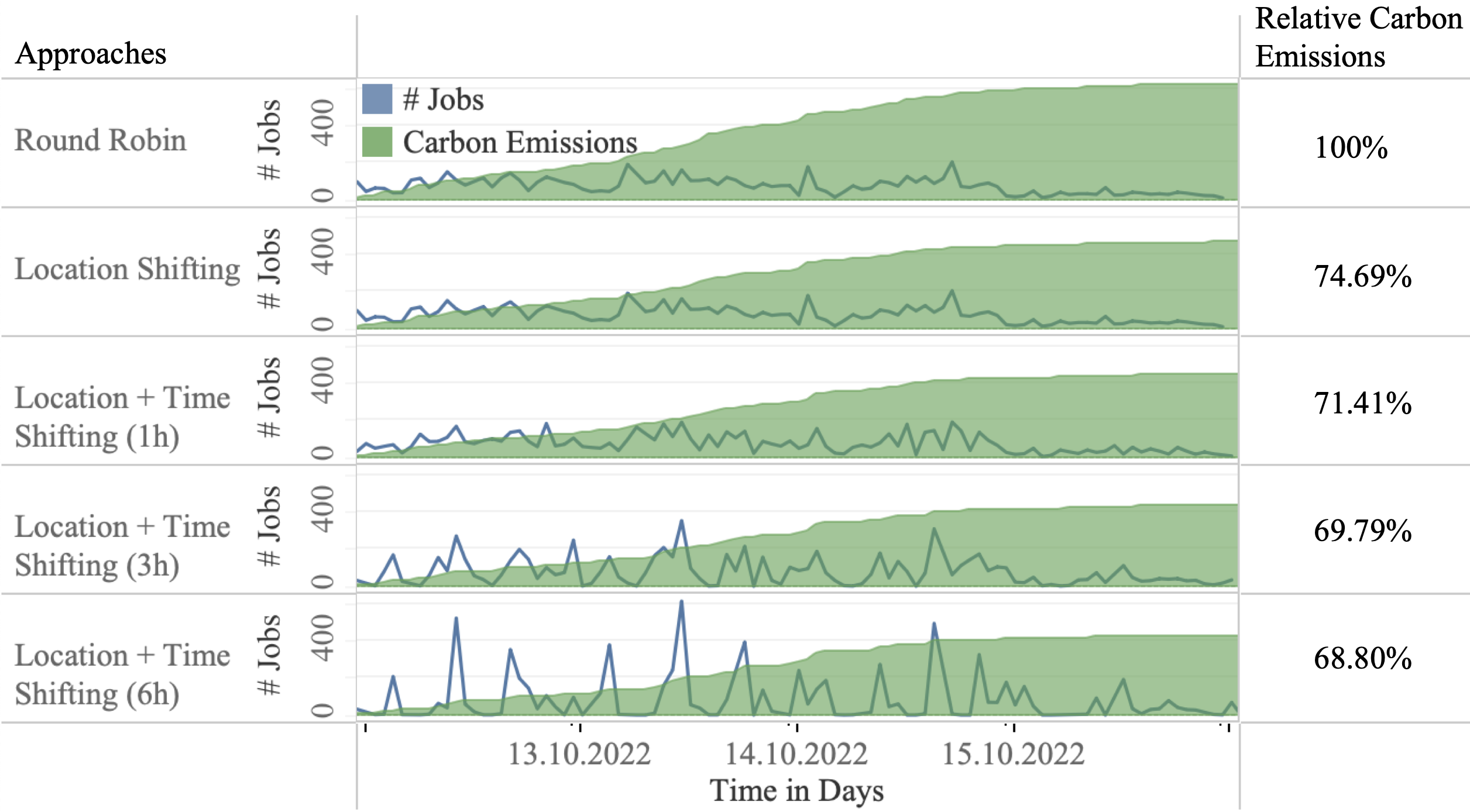}
 \caption{Amount of currently running jobs and accumulated carbon emissions for all five strategies.}
 \label{fig:carbonandjobs}
\end{figure*}

In Figure~\ref{fig:carbonandjobs}, it can be observed that in the round-robin baseline, the number of jobs is relatively evenly distributed, except for October 15th (Saturday), when the number of jobs is reduced over the weekend. Carbon emissions show some spikes on the afternoon of October 13th, as well as the early morning of October 12th and 14th. However, there are only a few periods where there are no or very low carbon emissions.

Since the \emph{Location Shifting} strategy involves no temporal scheduling, the distribution of jobs over time remained the same, but the carbon emissions in Figure~\ref{fig:carbonandjobs} are noticeable lower overall.
As a result, the overall carbon emissions could already be reduced by 25.31\%.  

In the \emph{Location$\,+\,$Time Shifting} strategy, the effect of additionally leveraging temporal flexibility becomes apparent.
For LTS-1, there was a noticeable reduction in carbon emissions, with some of the spikes shifting in time, which led to an improvement of 28.59\% compared to the baseline. In the cases of LTS-3 and LTS-6, the impact was even more pronounced: jobs were significantly shifted to periods with lower carbon intensity. %
This redistribution resulted in more concentrated job clusters, along with intermittent gaps that featured fewer or no jobs. 
The overall carbon emissions were reduced by 31.20\%.

\subsection{Limitations}
Although our data set used real-world data, we do not have the necessary information about data locality, the carbon cost of moving data, or resource utility.
Furthermore, power consumption was assumed to be steady, constant, and equal for each job -- meaning it only scales with its duration.
This assumption simplifies the evaluation but may not reflect real-world scenarios accurately.
Lastly, the data set also lacks information about sequential jobs and dependencies. Since our scheduler does not account for these cases, it would have been useful to evaluate dependency violations as well. This would provide insights into how our scheduler handles complex job dependencies and potential improvements needed in this area.
Therefore, our results are to be considered rather indicative, and results in real environments may be different.

\section{Conclusion and Future Work}
\label{sec:conclusion}
In this paper, we presented the idea of integrating approaches from carbon-aware computing into CI/CD services.
We discussed the opportunities and challenges when doing so and evaluated different carbon-aware scheduling strategies to demonstrate their possible effectiveness in carbon footprint reduction.

Solely relying on historical runtime data for estimating the job duration and
immediately allocating jobs to the region with the best window already resulted
in a 25.31\,\% reduction in overall carbon emissions. A key observation was that the
effective distribution of jobs across multiple regions with varying carbon intensities was a significant factor in reducing emissions. Building on this, our complete
scheduling approach introduces user input for estimated durations and deadlines, giving jobs the flexibility to be rescheduled in more carbon-efficient time
windows. When we employed this strategy, the improvement further increased
by an additional 3.28 -- 5.89\,\%, bringing the total improvement to approximately
31.2\,\%. It is important to clarify that our test data did not accurately represent
real-world user input; the estimated durations were closely aligned with actual
runtimes. Thus, we anticipate even greater improvements when accounting for
the inaccuracies commonly found in real user input.

As a next step, we will investigate the practical implications of different carbon-aware CI/CD scheduling by testing them on the internal CI/CD service of a large industrial company.
We anticipate that real-life deployments with global knowledge about the number and types of jobs offer additional opportunities as, for example, workflow duration and deadline estimates can be performed in a more sophisticated manner.
On the other hand, we expect new challenges in terms of data locality, resource constraints, and interdependencies between workflows which limit scheduling flexibility.
Lastly, we plan to simulate different seasons and datacenter locations through the use of a carbon-aware computing testbed~\cite{wiesner2023testbed} to give us a better picture of when and where there is substantial potential for savings.

\section*{Acknowledgments}

We sincerely thank WattTime for providing us with access to their marginal carbon intensity data.
This research was supported by the German Ministry for~Education and Research (BMBF) as Software Campus (grant 01IS17050).

\bibliographystyle{ieeetr}
\bibliography{literature}

\begin{thebibliography}{10}

\bibitem{masanet_recalibrating_2020}
E.~Masanet, A.~Shehabi, N.~Lei, S.~Smith, and J.~Koomey, ``Recalibrating global
  data center energy-use estimates,'' {\em Science}, vol.~367, no.~6481,
  pp.~984--986, 2020.

\bibitem{energy-efficient}
F.~Montevecchi, T.~Stickler, R.~Hintemann, and S.~Hinterholzer, {\em
  Energy-efficient {Cloud} {Computing} {Technologies} and {Policies} for an
  {Eco}-friendly {Cloud} {Market}. Final Study Report}.
\newblock LU: Publications Office of the European Union, 2020.

\bibitem{meyer_continuous_2014}
M.~Meyer, ``Continuous {Integration} and {Its} {Tools},'' {\em IEEE Software},
  vol.~31, no.~3, pp.~14--16, 2014.

\bibitem{cd_foundation_state_2022}
{CD Foundation}, ``State of {Continuous} {Delivery} {Report}: {The} {Evolution}
  of {Software} {Delivery} {Performance},'' 2022.

\bibitem{ibrahim2020depth}
M.~Ibrahim, S.~Nabi, A.~Baz, H.~Alhakami, M.~S. Raza, A.~Hussain, K.~Salah, and
  K.~Djemame, ``An in-depth empirical investigation of state-of-the-art
  scheduling approaches for cloud computing,'' {\em IEEE Access}, vol.~8,
  pp.~128282--128294, 2020.

\bibitem{WorldBank_CarbonPricing_2022}
W.~Bank, ``State and trends of carbon pricing 2022,'' tech. rep., Washington,
  DC: World Bank., 2022.

\bibitem{wiesner_lets_2021}
P.~Wiesner, I.~Behnke, D.~Scheinert, K.~Gontarska, and L.~Thamsen, ``Let's wait
  awhile: How temporal workload shifting can reduce carbon emissions in the
  cloud,'' in {\em ACM Middleware}, 2021.

\bibitem{Radovanovic_Google_2021}
A.~Radovanovic, R.~Koningstein, I.~Schneider, B.~Chen, A.~Duarte, B.~Roy,
  D.~Xiao, M.~Haridasan, P.~Hung, N.~Care, S.~Talukdar, E.~Mullen, K.~Smith,
  M.~Cottman, and W.~Cirne, ``Carbon-aware computing for datacenters,'' {\em
  IEEE Transactions on Power Systems}, 2022.

\bibitem{Fridgen_NotAllDoomAndGloom_2021}
G.~Fridgen, M.-F. K{\"o}rner, S.~Walters, and M.~Weibelzahl, ``Not all doom and
  gloom: How energy-intensive and temporally flexible data center applications
  may actually promote renewable energy sources,'' {\em Business {\&}
  Information Systems Engineering}, vol.~63, no.~3, 2021.

\bibitem{Zheng_MitigatingCurtailment_2020}
J.~Zheng, A.~A. Chien, and S.~Suh, ``Mitigating curtailment and carbon
  emissions through load migration between data centers,'' {\em Joule}, vol.~4,
  no.~10, 2020.

\bibitem{Zhou_CarbonAwareLoadBalancing_2013}
Z.~Zhou, F.~Liu, Y.~Xu, R.~Zou, H.~Xu, J.~C. Lui, and H.~Jin, ``Carbon-aware
  load balancing for geo-distributed cloud services,'' in {\em International
  Symposium on Modelling, Analysis and Simulation of Computer and
  Telecommunication Systems (MASCOTS)}, 2013.

\bibitem{moghaddam_carbon-aware_2015}
F.~Moghaddam, R.~Farrahi~Moghaddam, and M.~Cheriet, ``Carbon-aware distributed
  cloud: multi-level grouping genetic algorithm,'' {\em Cluster Computing},
  vol.~18, pp.~477--491, 2015.

\bibitem{hanafy2023carbonscaler}
W.~A. Hanafy, Q.~Liang, N.~Bashir, D.~Irwin, and P.~Shenoy, ``{CarbonScaler}:
  Leveraging cloud workload elasticity for optimizing carbon-efficiency,'' in
  {\em ACM SIGMETRICS / IFIP Performance}, 2024.

\bibitem{Lin_CouplingDatacentersPowerGrids_2021}
L.~Lin, V.~M. Zavala, and A.~Chien, ``Evaluating coupling models for cloud
  datacenters and power grids,'' in {\em ACM e-Energy}, 2021.

\bibitem{goiri_matching_2015}
I.~Goiri, M.~E. Haque, K.~Le, R.~Beauchea, T.~D. Nguyen, J.~Guitart, J.~Torres,
  and R.~Bianchini, ``Matching renewable energy supply and demand in green
  datacenters,'' {\em Ad Hoc Networks}, vol.~25, pp.~520--534, 2015.

\bibitem{Wiesner_Cucumber_2022}
P.~Wiesner, D.~Scheinert, T.~Wittkopp, L.~Thamsen, and O.~Kao, ``Cucumber:
  Renewable-aware admission control for delay-tolerant cloud and edge
  workloads,'' in {\em International European Conference on Parallel and
  Distributed Computing}, 2022.

\bibitem{abdalkareem_which_2021}
R.~Abdalkareem, S.~Mujahid, E.~Shihab, and J.~Rilling, ``Which {Commits} {Can}
  {Be} {CI} {Skipped}?,'' {\em IEEE Transactions on Software Engineering},
  vol.~47, no.~3, pp.~448--463, 2021.

\bibitem{ficher_assessing_2021}
M.~Ficher, F.~Berthoud, A.-L. Ligozat, P.~Sigonneau, M.~Wisslé, and
  B.~Tebbani, ``Assessing the carbon footprint of the data transmission on a
  backbone network,'' in {\em {Conference} on {Innovation} in {Clouds},
  {Internet} and {Networks} ({ICIN})}, 2021.

\bibitem{cirne_comprehensive_2001}
W.~Cirne and F.~Berman, ``A comprehensive model of the supercomputer
  workload,'' in {\em 4th {IEEE} {International} {Workshop} on {Workload}
  {Characterization}}, 2001.

\bibitem{tsafrir_backfilling_2007}
D.~Tsafrir, Y.~Etsion, and D.~G. Feitelson, ``Backfilling {Using}
  {System}-{Generated} {Predictions} {Rather} than {User} {Runtime}
  {Estimates},'' {\em IEEE Transactions on Parallel and Distributed Systems},
  vol.~18, no.~6, pp.~789--803, 2007.

\bibitem{tang_fault-aware_2009}
W.~Tang, Z.~Lan, N.~Desai, and D.~Buettner, ``Fault-aware, utility-based job
  scheduling on {Blue}, {Gene}/{P} systems,'' in {\em {IEEE} CLUSTER}, 2009.

\bibitem{ward_scheduling_2002}
W.~A. Ward, C.~L. Mahood, and J.~E. West, ``Scheduling {Jobs} on {Parallel}
  {Systems} {Using} a {Relaxed} {Backfill} {Strategy},'' in {\em Job
  {Scheduling} {Strategies} for {Parallel} {Processing}}, Springer, 2002.

\bibitem{wiesner2023testbed}
P.~Wiesner, I.~Behnke, and O.~Kao, ``A testbed for carbon-aware applications
  and systems,'' {\em arXiv:2306.09774 [cs.DC]}, 2023.

\end{thebibliography}
\end{document}